\documentclass[journal=acsnano,manuscript=article]{achemso}

\usepackage[version=3]{mhchem} 
\usepackage{multirow}
\usepackage{SIunits}
\usepackage{hyperref}
\usepackage{tabularx}
\usepackage[dvipsnames]{xcolor}
\usepackage[normalem]{ulem}
\usepackage{graphicx}
\usepackage{dcolumn}
\usepackage{amsmath}
\usepackage[english]{babel}
\usepackage[utf8]{inputenc}
\usepackage{stmaryrd}
\usepackage{mathtools}

\newcommand{\onlinecite}[1]{\hspace{-1 ex} \nocite{#1}\citenum{#1}}

\author{Amit Pawbake}
\email{amit.pawbake@lncmi.cnrs.fr}
\affiliation{LNCMI, UPR 3228, CNRS, EMFL, Université Grenoble Alpes, 38000 Grenoble, France}

\author{Sitaram Ramakrishnan}
\email{sitaram.ramakrishnan@neel.cnrs.fr}
\affiliation{Institut Néel CNRS/UGA UPR2940, 25 Rue des Martyrs, 38042 Grenoble, France}

\author{Kseniia Mosina}
\affiliation{Department of Inorganic Chemistry, University of Chemistry and Technology Prague,
Technicka 5, 166 28 Prague 6, Czech Republic}

\author{Aljoscha Soll}
\affiliation{Department of Inorganic Chemistry, University of Chemistry and Technology Prague,
Technicka 5, 166 28 Prague 6, Czech Republic}

\author{Zdenek Sofer}
\affiliation{Department of Inorganic Chemistry, University of Chemistry and Technology Prague,
Technicka 5, 166 28 Prague 6, Czech Republic}

\author{Pierre Rodiere}
\affiliation{Institut Néel CNRS/UGA UPR2940, 25 Rue des Martyrs, 38042 Grenoble, France}

\author{Rolf Heid}
\affiliation{Institute for Quantum Materials and Technologies, Karlsruhe Institute of Technology, 76131 Karlsruhe, Germany}

\author{Gaston Garbarino}
\affiliation{ESRF-European Synchrotron Radiation Facility, 38000 Cedex Grenoble, France}

\author{Pierre Bouvier}
\email{pierre.bouvier@neel.cnrs.fr}
\affiliation{Institut Néel CNRS/UGA UPR2940, 25 Rue des Martyrs, 38042 Grenoble, France}

\author{Clement Faugeras}
\email{clement.faugeras@lncmi.cnrs.fr}
\affiliation{LNCMI, UPR 3228, CNRS, EMFL, Université Grenoble Alpes, 38000 Grenoble, France}

\title[Pressure-induced symmetry breaking and polarity in the van der Waals magnet CrSBr]
  {Pressure-induced symmetry breaking and polarity in the van der Waals magnet CrSBr}

\begin{document}

\begin{tocentry}
\centering
\includegraphics[width=8cm,angle=0,clip]{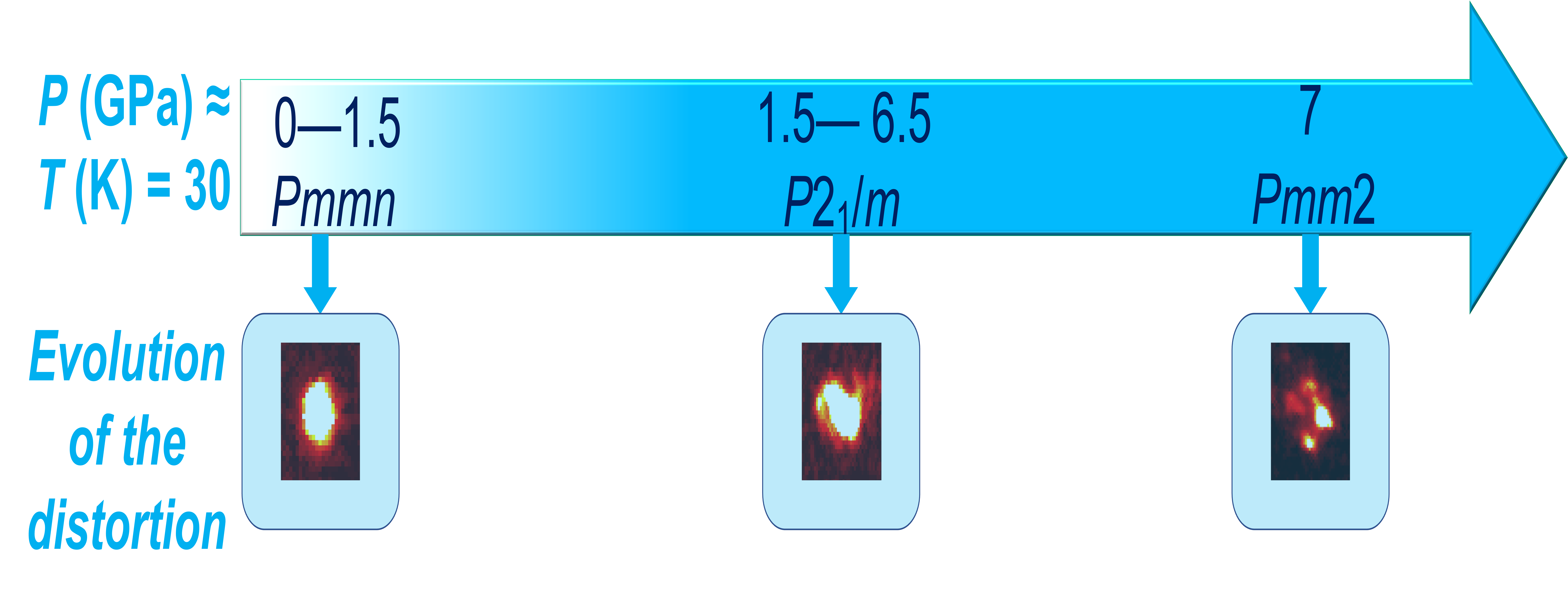}
\end{tocentry}

\begin{abstract}
The van der Waals magnet CrSBr has emerged as a fascinating platform for exploring coupled structural, magnetic, and electronic phenomena. Yet, the evolution of its crystal symmetry and electronic states under pressure and temperature remains unresolved. To elucidate this we employ low-temperature single-crystal X-ray diffraction, Raman scattering and photoluminescence measurements under hydrostatic pressures of up to 12 GPa, complemented by first-principles  calculations.We observe a cascade of pressure-induced structural transitions from the orthorhombic $Pmmn$ to monoclinic $P2_1/m$ phase at $\approx$ 1.5 GPa, followed by the emergence of a polar orthorhombic $Pmm$2 phase $\approx$ 6.5 GPa. The change in symmetry is accompanied by phonon anomalies and a direct-to-indirect bandgap crossover. Our results demonstrate pressure as a knob to engineer structure, symmetry and electronic properties in CrSBr in a coupled manner.
\end{abstract}

\date{\today }


\section{Introduction}
\label{sec:intro}
Symmetry breaking in quantum materials provides a powerful route to control emergent electronic and magnetic phenomena. Layered van der Waals (vdW) magnets serve as a quintessential example in this regard, as subtle structural distortions can strongly influence spin and electronic degrees of freedom~\cite{huang2017layer,coraux2025}. As interlayer coupling and crystal fields in these materials are sensitive to lattice spacing and symmetry, modest external perturbations such as strain~\cite{cenker2022reversible}, electric field~\cite{jiang2018controlling}, chemical pressure~\cite{telford2023designing,Badola2026} or hydrostatic pressure~\cite{pawbake2023magneto} can induce large changes in magnetic order, excitonic spectra, and electronic band topology. This tunability has made vdW magnets a focal point for both fundamental studies and device concepts in spintronics and magneto-optoelectronics~\cite{ziebel2024crsbr, tabataba2024doping, huang2018electrical}.

Chromium sulfur bromide (CrSBr) has emerged as a particularly attractive member of the vdW magnet family. Bulk CrSBr is an air-stable and semiconducting A-type antiferromagnet, with strong ferromagnetic intra-layer interactions and an antiferromagnetic interlayer coupling. It exhibits a Néel transition $T_{N}\sim132$~K and a direct optical bandgap in the near-infrared region~\cite{ziebel2024crsbr}. This material exhibits a remarkable coupling between its electronic band structure and magnetic ground state~\cite{wilson2021interlayer} allowing for the emission and detection of GHz magnons using visible light~\cite{Bae2022} and self-hybridized polaritons tunable by an external magnetic field~\cite{dirnberger2023magneto}. These properties make CrSBr an ideal candidate to investigate the interplay between lattice symmetry, magnetism, and optical response~\cite{lopez2022dynamic,pawbake2023raman,shao2025magnetically,Badola2026}.

Hydrostatic pressure is a powerful tuning parameter since it modifies interatomic distances and crystal symmetry without introducing chemical disorder. It can drive structural phase transitions that significantly alter magnetic exchange and electronic bands dispersions~\cite{song2019switching, li2019pressure}. Recent calculations and experimental studies have explored routes to tune CrSBr’s magnetic and electronic properties, including ligand substitution~\cite{telford2023designing,Badola2026} and uniaxial strain~\cite{cenker2022reversible}, and predicted that strain or compression should modify the interlayer exchange interaction and may strongly affect magnetic order under sufficient lattice deformation~\cite{esteras2022magnon,feng2025pressure}. Despite this interest, experimental work directly probing CrSBr under pressure at cryogenic temperature remains elusive~\cite{pawbake2023magneto}, with prior work limited to room-temperature studies~\cite{langston2025pressure}.

From single-crystal x-ray diffraction at 30 K under hydrostatic pressures up to 12 GPa, we resolve the pressure evolution of the crystal structure of CrSBr and unveil a sequence of structural transitions. The results corroborate low-temperature Raman spectroscopy that was used to probe lattice dynamics and photoluminescence (PL) measurements to investigate excitons as well as the evolution of optical properties as a function of pressure. We show that above $P=6.5$~GPa, CrSBr becomes an indirect band gap semiconductor. First-principles calculations were employed to model lattice dynamics in the different pressure-induced phases and to describe the evolution of the electronic band structure. Our results establish a subtle three-phase sequence, evolving from orthorhombic $Pmmn$ to  monoclinic $P2_1/m$ and ultimately rendering the crystal structure in a polar orthorhombic $Pmm2$ phase. These findings demonstrate that combining low temperature with hydrostatic pressure provides a unique pathway to induce polarity in CrSBr, thereby offering a compelling example of how subtle rearrangements of the crystal lattice can unlock profound changes in the magnetic order and electronic structure of a quantum material.

\section{RESULTS AND DISCUSSION}

\subsection{Pressure-Induced Structural Response of CrSBr at 30 K}

\begin{figure*}
\includegraphics[width=1\textwidth]{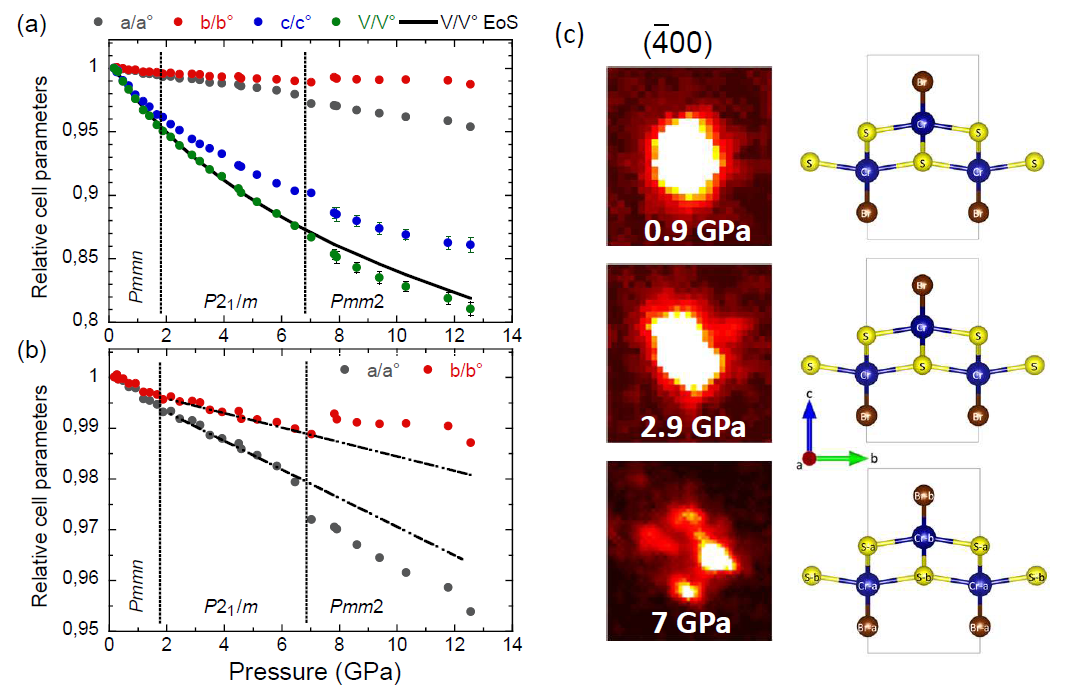}
\caption{\label{FigXRay} a) Normalized lattice parameters $a/a^o$, $b/b^o$, $c/c^o$ and volume $V/V^o$ obtained from structure refinement. $a^o=3.5103(1)$\AA{}, $b^o=4.7437(2)$\AA{}, $c^o=7.8320(7)$\AA{} and $V^o=130.41(11)$\AA{}$^3$ are taken at $p=0.19$~GPa. Full line $V/V^o$~EoS corresponds to the Birch-Murnaghan equation of state used below $P=6.5$~GPa.  b) Zoom on the normalized $a$ and $b$ lattice parameters showing the two phases transitions indicated by vertical dashed lines. c) (Left) Evolution of the (-400) Bragg reflection measured at three representative pressures at $T=30$~K showing the initial circular reflection at $p=0.9$~GPa, an elongated spot at $P=2.9$~GPa and the four spot reflection for $P>6.5$~GPa. (Right) Projection of the crystal structure of CrSBr onto the bc plane. The $P2_1/m$ phase differs from $Pmmn$ because the x-coordinate is no longer constrained to a special position, unlike in $Pmmn$. In the non-centrosymmetric Pmm2 phase the inversion center is lost and all atoms split into pairs of crystallographically distinct positions related by inversion symmetry. }
\end{figure*}

One fascinating aspect of van der Waals materials is how they deform when submitted to pressure. Being composed of covalently bound monolayers held together by weak van der Waals forces, they are strongly anisotropic and this is reflected in their deformation. Figure~\ref{FigXRay}a) displays the evolution of the lattice parameters as a function of the pressure showing, as for many van der Waals materials~\cite{Bykov2013,li2019pressure,Schaller2022,Schaller2023}, a strong deformation along the \textbf{c} axes and much weaker along the \textbf{a} or \textbf{b} axes. The unit cell volume is hence mainly imposed by the deformation along the \textbf{c} axes, a characteristic of the structural anisotropy of van der Waals materials. An abrupt change of the \textbf{c} axes parameter can be seen around $P=6.5$~GPa. At this pressure value, the lattice suddenly expands along the \textbf{b} axes and simultaneously contracts along the \textbf{a} axes. A zoom of the in-plane lattice parameters presented in Fig.~\ref{FigXRay}b), not only shows the $P=6.5$~GPa anomaly, but also another anomaly around $P=2$~GPa.

CrSBr undergoes a cascade of pressure-induced structural phase transitions at $30$~K upon compression up to $12$~GPa using He gas as pressure transmitting medium. However, the nature of these transitions is rather subtle, since the crystal structure of the material retains a semblance of orthorhombic $Pmmn$ symmetry, consistent with previous low-temperature XRD studies~\cite{lopez2022dynamic, telford2023designing}. To get valuable insights into these phase transitions, we trace the evolution of the ($\overline{4} 0 0$) Bragg reflection  under increasing pressure as presented in Figure \ref{FigXRay}c). The nearly circular reflection at $1$~GPa broadens and acquires an anisotropic shape above $2$~GPa, in line with the slope discontinuity observed near $2$~GPa in the $a$ and $b$ lattice parameters (see Fig.~\ref{FigXRay}b). At ambient pressure and low temperature, CrSBr has an orthorhombic $Pmmn$ symmetry which changes around $P=2$~GPa to a lower symmetry subgroup $P2_1/m$ with monoclinic symmetry (see supplemental material for details on the refinement procedure). The monoclinic distortion is rather negligible as reflected by the refined value of $\beta$ = $90.02(2)^\circ$, which further corroborates the nearly equal twin volume fractions of 0.5/0.5, emphasizing that the pressure-induced phase shows a near-orthorhombic $mmm$ symmetry. Pseudo-orthorhombic structures are ubiquitous across a wide range of materials including pressurized CrSBr, oxychlorides, ferroelectric AB$_2$X$_4$ (A = K, Rb, Cs; B = Zn, Co; X = Cl, Br, I) compounds such as Rb$_2$ZnCl$_4$~\cite{kotla2025a}, the antiferromagnet Eu$_2$Sb$_3$~\cite{chapuis1980a}, and the superconducting tri-layer nickelate La$_4$Ni$_3$O$_{10}$~\cite{ramakrishnan2026a}.

The partial overlap of twin domains, attributable to pseudo-merohedral twinning, evolves further under pressure. Reflections such as ($\overline{4} 0 0$) split into multiple components above $6.5$~GPa indicating a second phase transition. The second phase transition is supported by the poor agreement factors obtained for both the Pmmn and $P2_1/m$ models. Tables S5-S6 summarize our attempts to identify the high-pressure structure. The convergence of low agreement factors with a physically meaningful structural model provides compelling support for the assignment of polar orthorhombic $Pmm2$ symmetry, whereas the polar monoclinic Pn model was discarded because of its excessive number of parameters and non-physical atomic displacement parameters. This is an intriguing observation as orthorhombic symmetry is restored despite the concomitant loss of inversion symmetry. Table S2, S4 and S7  in the Supplemental material shows the atomic coordinates, where the deviation from the orthorhombic $Pmmn$ symmetry is minimal. The sequence of pressure induced structural phases in CrSBr at low temperature is summarized in Table~\ref{tab:CrSBr_crystalinfo} and details are outlined in the Supplemental material.

The unit cell volume follows a third-order Birch-Murnaghan equation of state (EoS) with parameters $V^{o}= 131.25(17)\AA{}^3$, $K^{o}= 27.1(1.3) GPa^{-1}$ and $K’=8.4(7)$ below $6.5$~GPa at $30K$. The first transition near $2$~GPa is second-order transition without any measurable discontinuity in the evolution of the unit cell volume with pressure. However, for the second transition at $P\sim6.5$~GPa, a noticeable jump in all the unit cell parameters and volume shows that this transition is most probably of first-order. In particular, as already mentioned, an increase in the \textbf{b} axes lattice parameter is observed, which also coincides with the magnetic easy axis. Considering the slope of the linear para- to antiferro-magnetic boundary $dT_{N}/dp = -12.6(1.0) K.GPa^{-1}$ proposed by Telford et al.~\cite{telford2023designing}, we expect to cross the same AFM to PM transition around 8.3(7) GPa at $30$~K. This critical pressure is in agreement with the value of 6.5 GPa  found for the $P2_{1}/m$ to \textit{Pmm}2 phase transition. This non-centrosymmetric structure was not expected as in general, the transition metal $d$ electrons, which are essential for magnetism, reduce the tendency for off-center ferroelectric distortion in oxygen octahedra~\cite{Hill2000}.

As inferred from Figure \ref{FigXRay}c, from a visual perspective the structural differences between these three phases are minimal. Nevertheless, they are quantitatively captured by the changes in atomic coordinates. In particular by the deviation of the atomic $x$-coordinates from the special positions upon the reduction of symmetry from $Pmmn$ to $P2_1/m$. Further compression drives the structure into the polar $Pmm2$ phase, where inversion symmetry is lost, as also mentioned in the caption of Figure \ref{FigXRay}. Additional details regarding the X-ray diffraction analysis and structural refinements are provided in the Supplemental Materials.

\begin{table*}[ht]
\caption{\label{tab:CrSBr_crystalinfo}%
Crystallographic data of CrSBr for three different phases at 30 K under pressure.}
\small
\centering
\begin{tabular}{cccc}
Pressure (GPa) & 0.92 & 2.9  & 7   \\
Crystal system & Orthorhombic  & Monoclinic  & Orthorhombic \\
Point symmetry & $mmm$ & 2/$m$ ($b$-unique) & $mm2$ \\
Space group (No)  & $Pmmn$~(59)  & $P$2$_{1}$/$m$~(11) &  $Pmm$2~(25)  \\
$\beta$ (deg) & 90  & 90.02(2) &  90 \\
No. of unique reflections (obs/all) & 76/83 & 118/139 & 100/131  \\
No. of parameters &13 & 20 & 26      \\
$R_{F }$  (obs/all) &0.0435/0.0449 &0.0543/0.0570 & 0.0453/0.0541   \\
\end{tabular}
\end{table*}

\subsection{High Pressure Raman scattering}

\begin{figure*}
\includegraphics[width=1\textwidth]{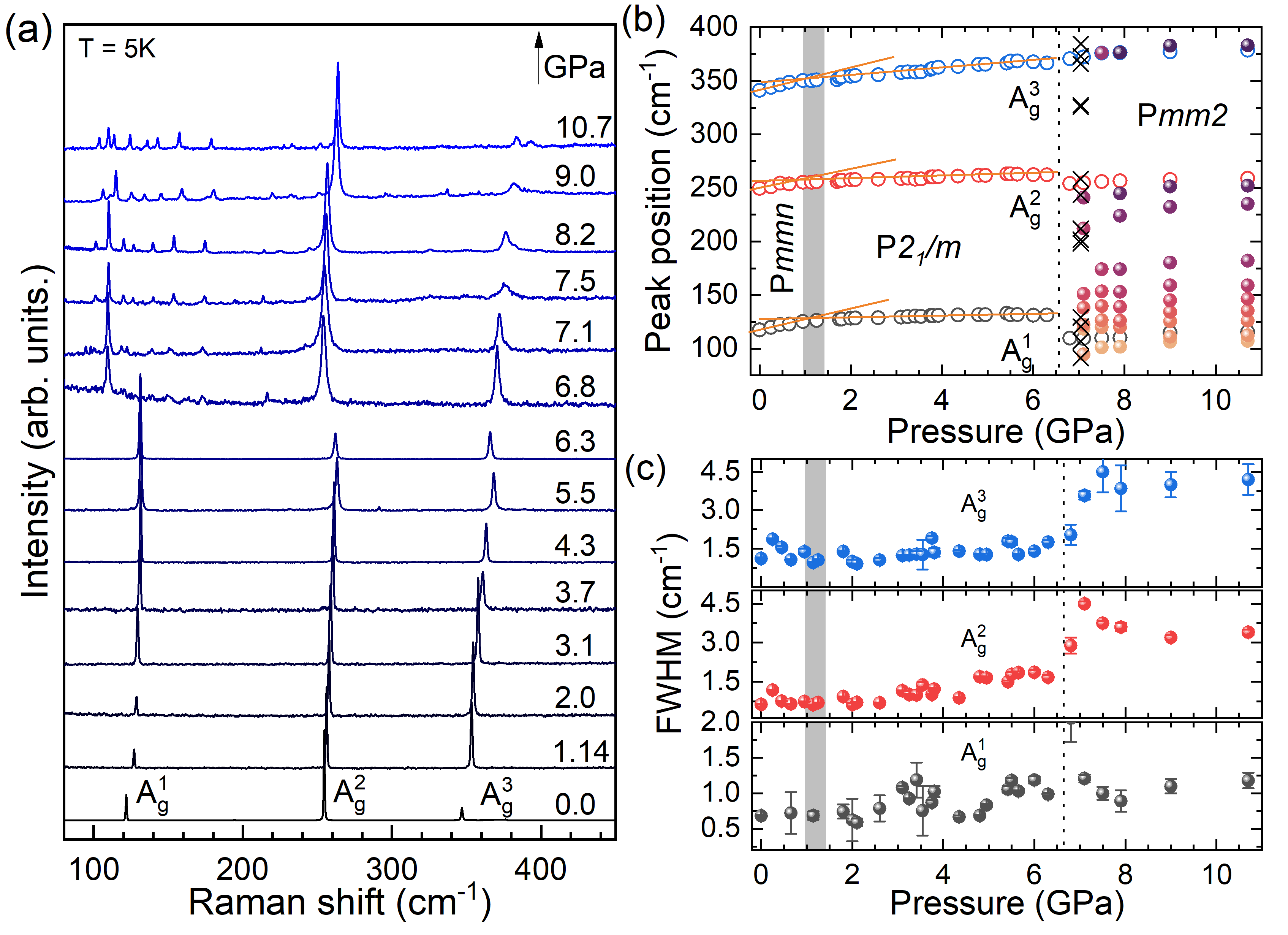}
\caption{\label{figRaman}
a) Unpolarized Raman scattering spectra of CrSBr as a function of pressure at 5~K. b) Evolution of the phonon energies ($A_g^{1-3}$).
Crosses mark the frequencies of the Raman active modes as obtained from DFT calculations for the $Pmm2$ relaxed structure. c) Full width at half maximum (FWHM) extracted by Lorentzian fitting as a function of pressure. The orange lines are guides to the eyes. Vertical dashed lines mark phase transitions at ~1.5 GPa and ~6.5 GPa. }
\end{figure*}

Figure~\ref{figRaman}a displays unpolarized Raman scattering spectra of CrSBr measured under pressures up to $11$~GPa at $5$~K using a 4:1 methanol-ethanol mixture as pressure transmitting medium. At ambient pressure, the spectrum is dominated by three prominent A$^{1,2,3}_{g}$ phonon modes, consistent with previous reports on CrSBr with an orthorhombic \textit{Pmmn} structure ($D_{2h}~mmm$ point group)~\cite{pawbake2023raman,lopez2022dynamic,Mondal2025,Sahu2025}. When applying pressure (Fig.~\ref{figRaman}b), these three phonon modes harden with an energy increasing linearly up to $P\sim1.5$~GPa. The phonons linewidths or lifetimes presented in Fig.~\ref{figRaman}c do not evolve in this pressure range. Above this pressure value and according to our X-ray analysis, CrSBr is in a monoclinic $P2_{1}/m$ phase. The Raman scattering response in this pressure induced phase is still composed of three prominent $A_g$ modes that harden with increasing pressure, but with a smaller rate. The spectrum is essentially similar to the one at ambient pressure with the same number of Raman active modes. We expect to observe 6A$_g$ in the monoclinic $P2_{1}/m$ sub-group however because the monoclinic distortion is almost negligible the intensity of the three additional A$_g^{4-6}$ modes is expected to be very low. We attribute the difference in the critical pressure values of these structural phase transitions determined from Raman scattering or XRD to the different temperature $5K$ and $30K$ chosen for the two experiments respectively.

The Raman scattering response is much more affected during the second structural transition identified by XRD, from the monoclinic $P2_{1}/m$ to the $Pmm2$ group which lacks inversion symmetry. In this group, one can expect fifteen modes 5A$_{1}$+5B$_{1}$+5B$_{2}$, with both Raman and infrared activity. Above $\sim6.5$~GPa, the Raman scattering response changes significantly with abrupt changes of the main phonon energies, in particular of phonons identified as  A$^{1,2}_{g}$, and with the appearance of 6 new phonon modes in the investigated range of energy, see Fig.~\ref{figRaman}a,b. The linewidth of phonon modes increases significantly in this high pressure phase. The simultaneous discontinuities in the pressure dependence of phonon energies, linewidths, and number of observed peaks thus provide a clear vibrational fingerprint of the high-pressure transition. In this phase, the phonon energies evolve smoothly with further compression. These changes indicate a doubling of the structural unit cell and enhanced lattice anharmonicity, consistent with the loss of inversion symmetry and the onset of the non-centrosymmetric \textit{Pmm2} phase identified by XRD. The Raman scattering response shows a continuous evolution above the critical pressure value identified by XRD. We believe that this observation reflects the coexistence of the two $P2_{1}/m$ and $Pmm2$ phases.

Using our DFT approach to calculate phonon energies in CrSBr at different pressures, see Table~S8, we show that the three main phonon modes of the $Pmmn$ phase do persist, with different symmetries, in the high pressure phases. This can be appreciated by the two following results: i) Experimental distortions away from the ideal $Pmmn$ are very tiny and thus, calculated frequencies for phonons are very close to the ones for ideal $Pmmn$, and ii) relaxation of the atomic positions drives the structure even closer to the ideal $Pmmn$ and calculated frequencies match that of the $Pmmn$ structure within numerical accuracy ($1$~cm$^{-1}$). In the Pmmn phase, there are 15 optical modes consisting of 9 Raman (3A$_{1g}$+3B$_{2g}$+3B$_{3g}$) and 6 IR modes (2B$_{1u}$+ 2B$_{2u}$+2B$_{3u}$). In the Pmm2 phase, one gets 5A$_{1}$+5B$_{1}$+5B$_{2}$, and all of them are Raman active. The relationship is A$_{1}$ $\rightarrow$ A$_{1g}$ + B$_{1u}$, B$_{1}$ $\rightarrow$ B$_{3g}$ + B$_{2u}$, and B$_{2}$ $\rightarrow$ B$_{2g}$ + B$_{3u}$. The new modes seen above 6.5 GPa in the Pmm2 phase could be coming either from the 3B$_{2g}$+3B$_{3g}$ Raman or from the 6 IR modes. We have calculated phonon energies in the high pressure $Pmm2$ phase of CrSBr, in the relaxed cell at P$=7.04$~GPa, see black crosses in Fig.~\ref{figRaman}b). The experimentally observed phonons largely correspond to the calculated ones. Taken together, the Raman data corroborate the structural sequence \textit{Pmmn} $\rightarrow$ \textit{P}2$_1$/\textit{m} $\rightarrow$ \textit{Pmm2} established by XRD.

\subsection{High pressure photoluminescence of CrSBr}

\begin{figure*}
\includegraphics[width=1\textwidth]{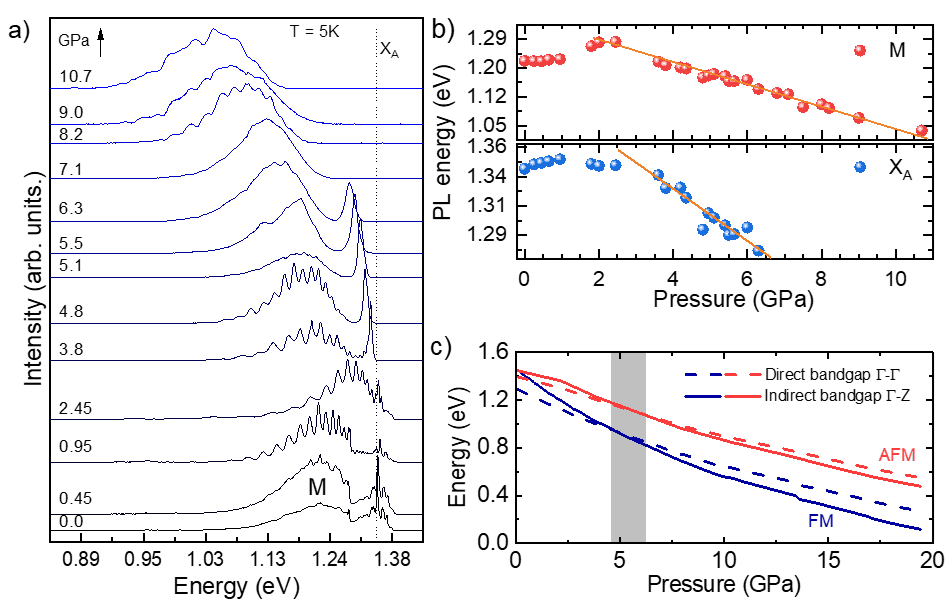}
\caption{\label{figPL}
 a) Pressure dependence of PL spectra of CrSBr at $T=5$~K measured up to 11~GPa in a broad energy range from 0.88 to 1.39~eV. b) Pressure evolution of the PL emission energies, revealing clear anomalies near $\sim$1.5~GPa and $\sim$6.5~GPa. c) DFT calculations of the pressure evolution of the direct band gap at $\Gamma$ (dashed line) and of the indirect bandgap (solid line) in the FM (blue) and AFM (red) states. The grey area indicates the direct to indirect band gap crossover. d,e) band structure of CrSBr in the FM state at $P=0.085$~GPa and at $P=10.33$~GPa , respectively. The black arrows represent the direct and indirect band gaps.}
\end{figure*}

The photoluminescence of CrSBr has recently attracted a lot of attention because it can probe the peculiar coupling between the electronic band structure and the magnetic state of the material~\cite{wilson2021interlayer}. Figure~\ref{figPL}a presents PL spectra of CrSBr measured at $T=5$~K as a function of pressure. At ambient pressure and at low temperature, this emission spectrum is composed of an exciton~\cite{wilson2021interlayer} and exciton-polariton~\cite{dirnberger2023magneto} related features, labeled $X_A$, close to $1.35$~eV, and of a large emission~\cite{pawbake2023magneto} centered at $1.2$~eV at ambient pressure and labeled M.

The pressure evolution of the PL of CrSBr is presented in Fig.~\ref{figPL}a-b. It is highly non-monotonic and shows distinct anomalies that correlate with the structural phase transitions identified by XRD and Raman scattering in the previous sections. Up to $\sim$2~GPa, all the PL emissions evolve smoothly: the $X_A$ peak shifts to higher energies while the energy of the broad band remains nearly constant. These monotonous evolutions suggest that the \textit{Pmmn} phase remains stable in this pressure range.

From $P\sim$2~GPa, the evolution of PL features changes drastically. In this pressure regime, the PL spectra are dominated by a sharp $X_A$ feature and a broad emission, both evolving smoothly and linearly toward lower energies up to $\sim$6.5~GPa. This pressure dependence is consistent with the pressure range of stability of the monoclinic \textit{P}2$_1$/\textit{m} phase. Up to this pressure, CrSBr retains a direct bandgap and an AFM ground state~\cite{pawbake2023magneto}. Nevertheless, the spectral shape of the $X_A$ line evolves: from a multicomponent feature, it evolves to a single, significantly broader, line, suggesting the disappearance of exciton-polariton spectral signatures. We interpret this evolution as resulting from the fact that the existence and observability of exciton-polariton depend on the oscillator strength of the exciton line and on its linewidth, both of which are affected by pressure. These observations are consistent with the results reported in Ref.~[\onlinecite{pawbake2023magneto}] showing that CrSBr at low temperature remains an AFM with pressure dependent magnetic properties.

Increasing further pressure, the most dramatic anomaly arises near $\sim6.5$~GPa, where the sharp $X_A$ emission corresponding to the exciton vanishes and only the broad band remains. This pressure value coincides with the appearance of new phonon modes in Raman scattering and with the phase transition to the non-centrosymmetric \textit{Pmm2} phase identified from XRD. Beyond this pressure, the remaining broad PL emission band redshifts linearly. This evolution suggests a profound change of the electronic properties of CrSBr under pressure, from a direct band gap semiconductor at ambient pressure in the $Pmmn$ phase and at moderate pressures up to $P\sim6.5$~GPa in the $P2_1/m$ monoclinic phase, to and indirect band gap semiconductor at higher pressure in the $Pmm2$ phase. The band gap of CrSBr remains indirect from $P=6.5$~GPa up to the highest pressure of $P=10.7$~GPa investigated in the work.

These interpretations are strongly supported by our DFT calculations of the electronic band structure of CrSBr and of the band gap presented in Fig.~\ref{figPL}c. At moderate pressure, CrSBr is a direct band gap semiconductor with a bright exciton at $1.32$~eV. The direct band gap is found at the $\Gamma$ point. When applying pressure, the maximum of the valence band changes from the $\Gamma$ to the $Z$ point and CrSBr turns to an indirect band gap semiconductor, as evidenced by the band structures presented in the Supplemental material. We can estimate the pressure value at which this direct to indirect band gap crossover occurs by calculating the pressure evolution of the $\Gamma$ to $\Gamma$ and of the $\Gamma$ to $Z$ band gaps in the ferromagnetic and in the antiferromagnetic states. Figure~\ref{figPL}c shows that independent of the magnetic ground state considered, the crossover occurs around $P=6$~GPa, in good agreement with the experimentally determined value of $P=6.3$~GPa. For pressures in the range $P=20-25$~GPa, CrSBr is expected to show a metallic behavior as suggested by our calculations showing a pronounced decrease of the band gap. The pressure-induced indirect bandgap in CrSBr could enable novel optoelectronic devices, where the exciton stability is tuned via hydrostatic pressure rather than chemical doping~\cite{telford2023designing,Badola2026}.

\section{\label{sec:level4}conclusion}

We have investigated the high-pressure behavior of the layered van der Waals magnet CrSBr up to 12~GPa at cryogenic temperatures using synchrotron XRD, Raman spectroscopy, PL, and first-principles calculations. At ambient pressure, CrSBr adopts an orthorhombic \textit{P}mmn structure with A-type AFM ground state. XRD measurements reveal successive pressure-driven structural transitions, $Pmmn~\rightarrow~P2_1/m~\rightarrow~Pmm2$, occurring near $1.5$ and $6.5$~GPa, respectively. Raman spectroscopy detects phonon slope discontinuities and the appearance of additional modes above $\sim$6.5~GPa, consistent with symmetry lowering in the high-pressure phase. PL measurements identify two critical anomalies at $\sim$2 and 6.5~GPa that correlate with the structural transitions and suggests a transformation from a direct to an indirect bandgap semiconductor above $\sim$6.5~GPa. Together, these results establish CrSBr as a model two-dimensional magnet in which pressure provides a powerful tuning knob to manipulate structure, symmetry, and electronic properties in a coupled manner. Our findings demonstrate that hydrostatic pressure can reversibly tune CrSBr through three distinct structural phases with distinct phononic and optical signatures and reveal a previously unreported polar phase. This suggests CrSBr as a model system for pressure-controlled spintronics and optoelectronics, with potential applications in dynamic magnetic and excitonic devices.

\section{Experimental and computational Details}
\label{sec:details}

\subsection{Synthesis of bulk CrSBr single crystals}
CrSBr single crystals were synthesized using a chemical vapor transport (CVT) method. As already reported~\cite{pawbake2023raman, pawbake2023magneto, long2023intrinsic}, chromium, sulfur, and bromine in the stoichiometric ratio of CrSBr were added and sealed in a quartz tube under a high vacuum. We use a two zone furnace to allow for a temperature gradient within the tube and  the coexistence of a hot region for the sublimation of the various elements, and of a cold zone for the deposition and growth of bulk CrSBr. The synthesized crystals have then been characterized by XRD and Raman scattering.

\subsection{High pressure Synchrotron XRD at 30K}
Single crystal X-ray diffraction (XRD) experiment were done at ID15B beamline (ESRF Grenoble) with a monochromatic wavelength $\lambda =0.41020$\AA{} and a $1\times 2 \mu$m focused beam. Diffraction images were collected using a Eiger2 CdTe 9M pixel counter detector (DECTRIS AG, Baden-Daettwil, Switzerland) in shutterless mode and continuous rotation of the DAC around the vertical $\omega$ axis in a range ±32°, with an angular step of $\Delta\omega$=0.5° and an exposure time of 0.5 s/frame. The CrysAlisPro software package \cite{crysalis} was used for the analysis of the single-crystal XRD data (indexing, data integration, frame scaling, and absorption correction). A single crystal of Vanadinite [Pb$_5$(VO$_4$)$_3$Cl, $Pbca$ space group, $a = 8.8117(2)$ \AA{}, $b = 5.18320(10)$ \AA{}, and $c = 18.2391(3)$ \AA{}] was used to calibrate the instrumental model in the CrysAlisPro software, i.e., sample-detector distance, detector’s origin, goniometer angles offsets, rotation of both the X-ray beam and detector around the instrument axis.
High-pressure experiments at low temperature were performed using membrane-driven diamond-anvil cells (MDACs) with 450 $\mu$m bevelled diamond culets, mounted on a home-built helium flow cryostat. Rhenium gasket were used for the pressure chambers. The pressure transmitting medium used was Helium, loaded at 1.2 kbar, to ensure high hydostaticity conditions up to the highest pressure reached in this study. The pressure was measured using the R1-line emission of a ruby ball placed close to the sample using the International Practical Pressure Scale IPPS-Ruby2020 equation of state \cite{Shen2020}.

\subsection{Low temperature and high pressure optical spectroscopy}
Pressure have been applied using a symmetric diamond-anvil cell (DACs) with culets of 500 $\mu$m. Inconel gasket were used for the pressure chambers. The pressure transmitting medium was a 4:1 methanol-ethanol mixture and the pressure was determined with the luminescence of a ruby ball crystal inserted in the pressure chamber. Raman and photoluminescence (PL) spectroscopy were performed at cryogenic temperatures (down to 5 K) using a home-built free beam setup in backscattering integrated with a Helium bath cryostat and superconducting magnet, as described in detail in our previous instrumentation work~\cite{breslavetz2021spatially} In this configuration, the same optical path allows acquisition of high-resolution Raman spectra and magneto-PL under applied fields up to 14 Tesla, and a magnetic field applied along the \textbf{b} easy axis while maintaining precise control of temperature and pressure. All Raman spectroscopic measurements were performed using a 514.5 nm laser excitation in combination with a commercial long-working-distance objective (working distance: 12 mm and numerical aperture: 0.35). The laser power was kept below $300$~$\mu$W. Raman scattering spectra were recorded with a 2000 gr/mm grating, providing a spectral resolution better than 0.3~cm$^{-1}$. A liquid-nitrogen-cooled Pylon CCD detector was used for data acquisition. Photoluminescence (PL) measurements were conducted using the same objective and excitation laser. For PL collection, an InGaAs camera was employed to detect a broad energy range of PL signals, with the excitation power maintained at 10~\textmu W.

\subsection{Computational details}

Density-functional theory (DFT) based calculations of electronic structure and lattice dynamical properties were performed using the same approach as outlined in previous publications \cite{pawbake2023raman,Badola2026}. In the following, we briefly recapitulate the main aspects of this approach. The electronic structure was obtained using the mixed-basis pseudopotential method \cite{meyer}, which employs an expansion of the valence states in a combination of plane waves and local functions at atomic sites to efficiently describe more localized components of the valence states. Spin-polarized calculations were carried out using the general-gradient approximation \cite{pbe1996}. Correlation corrections for the $d$-orbitals of Cr were incorporated within the DFT+U scheme, using $U=4$~eV and $J=1$~eV.

To analyze the pressure dependence of the gap, band structure calculations were performed for both the A-type AFM order with doubled unit cell, as well as the FM order, which can be described with the primitive unit cell. Pressure was simulated by using the experimental lattice constants obtained from the high-pressure synchrotron XRD refinements, while the atomic positions were relaxed to get a force-free structure. Gap values were taken from the Kohn-Sham band structure without further corrections.

Lattice dynamical properties were calculated within the linear response or density-functional
perturbation theory (DFPT) implemented in the mixed-basis pseudopotential method \cite{heid1999}. To study the changes in the Raman spectrum across the structural phase transitions, this technique was applied to both the high-symmetry $Pmmn$ and the lower symmetry $P2_1$/m and $Pmm2$ space groups using the experimentally determined cell parameters (lattice constants and atomic positions) from XRD refinements for P=2.47~GPa and P=7.04~GPa, respectively. In addition, we also preformed phonon calculations for structures with relaxed atomic positions, i.e. for force-free configurations.

\begin{acknowledgement}
This work was partially supported by LNCMI, a member of the European Magnetic Field Laboratory (EMFL). This work is also partially supported by the France 2030 government investment plan, managed by the French National Research Agency under Grant Reference PEPR SPIN-SPINMAT ANR-22-EXSP-0007, by ANR-23-QUAC-0004, and by CEFIPRA CSRP Project No.~7104-2. We acknowledge ESRF for provision of synchrotron radiation facilities under proposals MA-6379 and MA-6597. S.R. thanks the support from the Agence Nationale de la Recherche under the project SUPERNICKEL (Grant No.ANR-21-CE30-0041-04) and also the European Research Council (ERC) under the European Union's Horizon 2020 research and innovation program (Grant Agreement No. 865826). Z.S.\ and K.M.\ were supported by the ERC-CZ program (project LL2101) from the Ministry of Education, Youth and Sports (MEYS) and by the project Advanced Functional Nanorobots (reg.~No.~CZ.02.1.01/0.0/0.0/15-003/0000444), financed by the ERDF). A.S. was supported by the Czech Science Foundation Grant No. 24-11465S.
R.H. acknowledges support by the state of Baden-W\"{u}rttemberg through bwHPC.
\end{acknowledgement}

\begin{suppinfo}
Additional data and analysis are available in the supporting information.
They include the detailed XRD analysis leading to the identification of the high pressure phases. We also present the results of our band structure calculation showing the direct to indirect band gap crossover, and the result of phonon calculations considering the $Pmmn$ group, the experimentally determined cell parameters or the relaxed cell parameters.
\end{suppinfo}


\begin{mcitethebibliography}{38}
\providecommand*\natexlab[1]{#1}
\providecommand*\mciteSetBstSublistMode[1]{}
\providecommand*\mciteSetBstMaxWidthForm[2]{}
\providecommand*\mciteBstWouldAddEndPuncttrue
  {\def\EndOfBibitem{\unskip.}}
\providecommand*\mciteBstWouldAddEndPunctfalse
  {\let\EndOfBibitem\relax}
\providecommand*\mciteSetBstMidEndSepPunct[3]{}
\providecommand*\mciteSetBstSublistLabelBeginEnd[3]{}
\providecommand*\EndOfBibitem{}
\mciteSetBstSublistMode{f}
\mciteSetBstMaxWidthForm{subitem}{(\alph{mcitesubitemcount})}
\mciteSetBstSublistLabelBeginEnd
  {\mcitemaxwidthsubitemform\space}
  {\relax}
  {\relax}

\bibitem[Huang \latin{et~al.}(2017)Huang, Clark, Navarro-Moratalla, Klein,
  Cheng, Seyler, Zhong, Schmidgall, McGuire, Cobden, \latin{et~al.}
  others]{huang2017layer}
Huang,~B.; Clark,~G.; Navarro-Moratalla,~E.; Klein,~D.~R.; Cheng,~R.;
  Seyler,~K.~L.; Zhong,~D.; Schmidgall,~E.; McGuire,~M.~A.; Cobden,~D.~H.;
  others Layer-dependent ferromagnetism in a van der Waals crystal down to the
  monolayer limit. \emph{Nature} \textbf{2017}, \emph{546}, 270--273\relax
\mciteBstWouldAddEndPuncttrue
\mciteSetBstMidEndSepPunct{\mcitedefaultmidpunct}
{\mcitedefaultendpunct}{\mcitedefaultseppunct}\relax
\EndOfBibitem
\bibitem[Coraux \latin{et~al.}(2025)Coraux, Rougemaille, Robert, Faugeras,
  Saul, Grémaud, Hueso, Casanova, and Manchon]{coraux2025}
Coraux,~J.; Rougemaille,~N.; Robert,~C.; Faugeras,~C.; Saul,~A.; Grémaud,~B.;
  Hueso,~L.; Casanova,~F.; Manchon,~A. Colloquium: A critique on van der Waals
  and two-dimensional magnets. 2025;
  \url{https://arxiv.org/abs/2512.19417}\relax
\mciteBstWouldAddEndPuncttrue
\mciteSetBstMidEndSepPunct{\mcitedefaultmidpunct}
{\mcitedefaultendpunct}{\mcitedefaultseppunct}\relax
\EndOfBibitem
\bibitem[Cenker \latin{et~al.}(2022)Cenker, Sivakumar, Xie, Miller, Thijssen,
  Liu, Dismukes, Fonseca, Anderson, Zhu, \latin{et~al.}
  others]{cenker2022reversible}
Cenker,~J.; Sivakumar,~S.; Xie,~K.; Miller,~A.; Thijssen,~P.; Liu,~Z.;
  Dismukes,~A.; Fonseca,~J.; Anderson,~E.; Zhu,~X.; others Reversible
  strain-induced magnetic phase transition in a van der Waals magnet.
  \emph{Nature Nanotechnology} \textbf{2022}, \emph{17}, 256--261\relax
\mciteBstWouldAddEndPuncttrue
\mciteSetBstMidEndSepPunct{\mcitedefaultmidpunct}
{\mcitedefaultendpunct}{\mcitedefaultseppunct}\relax
\EndOfBibitem
\bibitem[Jiang \latin{et~al.}(2018)Jiang, Li, Wang, Mak, and
  Shan]{jiang2018controlling}
Jiang,~S.; Li,~L.; Wang,~Z.; Mak,~K.~F.; Shan,~J. Controlling magnetism in 2D
  CrI3 by electrostatic doping. \emph{Nature nanotechnology} \textbf{2018},
  \emph{13}, 549--553\relax
\mciteBstWouldAddEndPuncttrue
\mciteSetBstMidEndSepPunct{\mcitedefaultmidpunct}
{\mcitedefaultendpunct}{\mcitedefaultseppunct}\relax
\EndOfBibitem
\bibitem[Telford \latin{et~al.}(2023)Telford, Chica, Ziebel, Xie, Manganaro,
  Huang, Cox, Dismukes, Zhu, Walsh, \latin{et~al.}
  others]{telford2023designing}
Telford,~E.~J.; Chica,~D.~G.; Ziebel,~M.~E.; Xie,~K.; Manganaro,~N.~S.;
  Huang,~C.-Y.; Cox,~J.; Dismukes,~A.~H.; Zhu,~X.; Walsh,~J.~P.; others
  Designing magnetic properties in CrSBr through hydrostatic pressure and
  ligand substitution. \emph{Advanced Physics Research} \textbf{2023},
  \emph{2}, 2300036\relax
\mciteBstWouldAddEndPuncttrue
\mciteSetBstMidEndSepPunct{\mcitedefaultmidpunct}
{\mcitedefaultendpunct}{\mcitedefaultseppunct}\relax
\EndOfBibitem
\bibitem[Badola \latin{et~al.}(2026)Badola, Pawbake, Wu, S\"{o}ll, Sofer, Heid,
  and Faugeras]{Badola2026}
Badola,~S.; Pawbake,~A.; Wu,~B.; S\"{o}ll,~A.; Sofer,~Z.; Heid,~R.;
  Faugeras,~C. van der Waals CrSBr Alloys with Tunable Magnetic and Optical
  Properties. \emph{Nano Letters} \textbf{2026}, \emph{26}, 773\relax
\mciteBstWouldAddEndPunctfalse
\mciteSetBstMidEndSepPunct{\mcitedefaultmidpunct}
{}{\mcitedefaultseppunct}\relax
\EndOfBibitem
\bibitem[Pawbake \latin{et~al.}(2023)Pawbake, Pelini, Mohelsky, Jana,
  Breslavetz, Cho, Orlita, Potemski, Measson, Wilson, \latin{et~al.}
  others]{pawbake2023magneto}
Pawbake,~A.; Pelini,~T.; Mohelsky,~I.; Jana,~D.; Breslavetz,~I.; Cho,~C.-W.;
  Orlita,~M.; Potemski,~M.; Measson,~M.-A.; Wilson,~N.~P.; others
  Magneto-optical sensing of the pressure driven magnetic ground states in bulk
  CrSBr. \emph{Nano Letters} \textbf{2023}, \emph{23}, 9587--9593\relax
\mciteBstWouldAddEndPuncttrue
\mciteSetBstMidEndSepPunct{\mcitedefaultmidpunct}
{\mcitedefaultendpunct}{\mcitedefaultseppunct}\relax
\EndOfBibitem
\bibitem[Ziebel \latin{et~al.}(2024)Ziebel, Feuer, Cox, Zhu, Dean, and
  Roy]{ziebel2024crsbr}
Ziebel,~M.~E.; Feuer,~M.~L.; Cox,~J.; Zhu,~X.; Dean,~C.~R.; Roy,~X. CrSBr: an
  air-stable, two-dimensional magnetic semiconductor. \emph{Nano letters}
  \textbf{2024}, \emph{24}, 4319--4329\relax
\mciteBstWouldAddEndPuncttrue
\mciteSetBstMidEndSepPunct{\mcitedefaultmidpunct}
{\mcitedefaultendpunct}{\mcitedefaultseppunct}\relax
\EndOfBibitem
\bibitem[Tabataba-Vakili \latin{et~al.}(2024)Tabataba-Vakili, Nguyen, Rupp,
  Mosina, Papavasileiou, Watanabe, Taniguchi, Maletinsky, Glazov, Sofer,
  \latin{et~al.} others]{tabataba2024doping}
Tabataba-Vakili,~F.; Nguyen,~H.~P.; Rupp,~A.; Mosina,~K.; Papavasileiou,~A.;
  Watanabe,~K.; Taniguchi,~T.; Maletinsky,~P.; Glazov,~M.~M.; Sofer,~Z.; others
  Doping-control of excitons and magnetism in few-layer CrSBr. \emph{Nature
  Communications} \textbf{2024}, \emph{15}, 4735\relax
\mciteBstWouldAddEndPuncttrue
\mciteSetBstMidEndSepPunct{\mcitedefaultmidpunct}
{\mcitedefaultendpunct}{\mcitedefaultseppunct}\relax
\EndOfBibitem
\bibitem[Huang \latin{et~al.}(2018)Huang, Clark, Klein, MacNeill,
  Navarro-Moratalla, Seyler, Wilson, McGuire, Cobden, Xiao, \latin{et~al.}
  others]{huang2018electrical}
Huang,~B.; Clark,~G.; Klein,~D.~R.; MacNeill,~D.; Navarro-Moratalla,~E.;
  Seyler,~K.~L.; Wilson,~N.; McGuire,~M.~A.; Cobden,~D.~H.; Xiao,~D.; others
  Electrical control of 2D magnetism in bilayer CrI 3. \emph{Nature
  nanotechnology} \textbf{2018}, \emph{13}, 544--548\relax
\mciteBstWouldAddEndPuncttrue
\mciteSetBstMidEndSepPunct{\mcitedefaultmidpunct}
{\mcitedefaultendpunct}{\mcitedefaultseppunct}\relax
\EndOfBibitem
\bibitem[Wilson \latin{et~al.}(2021)Wilson, Lee, Cenker, Xie, Dismukes,
  Telford, Fonseca, Sivakumar, Dean, Cao, \latin{et~al.}
  others]{wilson2021interlayer}
Wilson,~N.~P.; Lee,~K.; Cenker,~J.; Xie,~K.; Dismukes,~A.~H.; Telford,~E.~J.;
  Fonseca,~J.; Sivakumar,~S.; Dean,~C.; Cao,~T.; others Interlayer electronic
  coupling on demand in a 2D magnetic semiconductor. \emph{Nature Materials}
  \textbf{2021}, \emph{20}, 1657--1662\relax
\mciteBstWouldAddEndPuncttrue
\mciteSetBstMidEndSepPunct{\mcitedefaultmidpunct}
{\mcitedefaultendpunct}{\mcitedefaultseppunct}\relax
\EndOfBibitem
\bibitem[Bae \latin{et~al.}(2022)Bae, Wang, Scheie, Xu, Chica, Diederich,
  Cenker, Ziebel, Bai, Ren, Dean, Delor, Xu, Roy, Kent, and Zhu]{Bae2022}
Bae,~Y.~J. \latin{et~al.}  Exciton-coupled coherent magnons in a 2D
  semiconductor. \emph{Nature} \textbf{2022}, \emph{609}, 282–286\relax
\mciteBstWouldAddEndPuncttrue
\mciteSetBstMidEndSepPunct{\mcitedefaultmidpunct}
{\mcitedefaultendpunct}{\mcitedefaultseppunct}\relax
\EndOfBibitem
\bibitem[Dirnberger \latin{et~al.}(2023)Dirnberger, Quan, Bushati, Diederich,
  Florian, Klein, Mosina, Sofer, Xu, Kamra, \latin{et~al.}
  others]{dirnberger2023magneto}
Dirnberger,~F.; Quan,~J.; Bushati,~R.; Diederich,~G.~M.; Florian,~M.;
  Klein,~J.; Mosina,~K.; Sofer,~Z.; Xu,~X.; Kamra,~A.; others Magneto-optics in
  a van der Waals magnet tuned by self-hybridized polaritons. \emph{Nature}
  \textbf{2023}, \emph{620}, 533--537\relax
\mciteBstWouldAddEndPuncttrue
\mciteSetBstMidEndSepPunct{\mcitedefaultmidpunct}
{\mcitedefaultendpunct}{\mcitedefaultseppunct}\relax
\EndOfBibitem
\bibitem[L{\'o}pez-Paz \latin{et~al.}(2022)L{\'o}pez-Paz, Guguchia,
  Pomjakushin, Witteveen, Cervellino, Luetkens, Casati, Morpurgo, and von
  Rohr]{lopez2022dynamic}
L{\'o}pez-Paz,~S.~A.; Guguchia,~Z.; Pomjakushin,~V.~Y.; Witteveen,~C.;
  Cervellino,~A.; Luetkens,~H.; Casati,~N.; Morpurgo,~A.~F.; von Rohr,~F.~O.
  Dynamic magnetic crossover at the origin of the hidden-order in van der Waals
  antiferromagnet CrSBr. \emph{Nature Communications} \textbf{2022}, \emph{13},
  4745\relax
\mciteBstWouldAddEndPuncttrue
\mciteSetBstMidEndSepPunct{\mcitedefaultmidpunct}
{\mcitedefaultendpunct}{\mcitedefaultseppunct}\relax
\EndOfBibitem
\bibitem[Pawbake \latin{et~al.}(2023)Pawbake, Pelini, Wilson, Mosina, Sofer,
  Heid, and Faugeras]{pawbake2023raman}
Pawbake,~A.; Pelini,~T.; Wilson,~N.~P.; Mosina,~K.; Sofer,~Z.; Heid,~R.;
  Faugeras,~C. Raman scattering signatures of strong spin-phonon coupling in
  the bulk magnetic van der Waals material CrSBr. \emph{Physical Review B}
  \textbf{2023}, \emph{107}, 075421\relax
\mciteBstWouldAddEndPuncttrue
\mciteSetBstMidEndSepPunct{\mcitedefaultmidpunct}
{\mcitedefaultendpunct}{\mcitedefaultseppunct}\relax
\EndOfBibitem
\bibitem[Shao \latin{et~al.}(2025)Shao, Dirnberger, Qiu, Acharya, Terres,
  Telford, Pashov, Kim, Ruta, Chica, \latin{et~al.}
  others]{shao2025magnetically}
Shao,~Y.; Dirnberger,~F.; Qiu,~S.; Acharya,~S.; Terres,~S.; Telford,~E.~J.;
  Pashov,~D.; Kim,~B.~S.; Ruta,~F.~L.; Chica,~D.~G.; others Magnetically
  confined surface and bulk excitons in a layered antiferromagnet. \emph{Nature
  Materials} \textbf{2025}, 1--8\relax
\mciteBstWouldAddEndPuncttrue
\mciteSetBstMidEndSepPunct{\mcitedefaultmidpunct}
{\mcitedefaultendpunct}{\mcitedefaultseppunct}\relax
\EndOfBibitem
\bibitem[Song \latin{et~al.}(2019)Song, Fei, Yankowitz, Lin, Jiang, Hwangbo,
  Zhang, Sun, Taniguchi, Watanabe, \latin{et~al.} others]{song2019switching}
Song,~T.; Fei,~Z.; Yankowitz,~M.; Lin,~Z.; Jiang,~Q.; Hwangbo,~K.; Zhang,~Q.;
  Sun,~B.; Taniguchi,~T.; Watanabe,~K.; others Switching 2D magnetic states via
  pressure tuning of layer stacking. \emph{Nature materials} \textbf{2019},
  \emph{18}, 1298--1302\relax
\mciteBstWouldAddEndPuncttrue
\mciteSetBstMidEndSepPunct{\mcitedefaultmidpunct}
{\mcitedefaultendpunct}{\mcitedefaultseppunct}\relax
\EndOfBibitem
\bibitem[Li \latin{et~al.}(2019)Li, Jiang, Sivadas, Wang, Xu, Weber,
  Goldberger, Watanabe, Taniguchi, Fennie, \latin{et~al.}
  others]{li2019pressure}
Li,~T.; Jiang,~S.; Sivadas,~N.; Wang,~Z.; Xu,~Y.; Weber,~D.; Goldberger,~J.~E.;
  Watanabe,~K.; Taniguchi,~T.; Fennie,~C.~J.; others Pressure-controlled
  interlayer magnetism in atomically thin CrI3. \emph{Nature materials}
  \textbf{2019}, \emph{18}, 1303--1308\relax
\mciteBstWouldAddEndPuncttrue
\mciteSetBstMidEndSepPunct{\mcitedefaultmidpunct}
{\mcitedefaultendpunct}{\mcitedefaultseppunct}\relax
\EndOfBibitem
\bibitem[Esteras \latin{et~al.}(2022)Esteras, Rybakov, Ruiz, and
  Baldov{\'\i}]{esteras2022magnon}
Esteras,~D.~L.; Rybakov,~A.; Ruiz,~A.~M.; Baldov{\'\i},~J.~J. Magnon
  straintronics in the 2D van der Waals ferromagnet CrSBr from
  first-principles. \emph{Nano Letters} \textbf{2022}, \emph{22},
  8771--8778\relax
\mciteBstWouldAddEndPuncttrue
\mciteSetBstMidEndSepPunct{\mcitedefaultmidpunct}
{\mcitedefaultendpunct}{\mcitedefaultseppunct}\relax
\EndOfBibitem
\bibitem[Feng \latin{et~al.}(2025)Feng, Fu, Lu, Ke, and
  L{\"u}]{feng2025pressure}
Feng,~Y.; Fu,~W.; Lu,~Q.; Ke,~S.-S.; L{\"u},~H.-F. Pressure induced magnetic
  transition and metallization in antiferromagnet CrSBr bilayer.
  \emph{Computational Materials Science} \textbf{2025}, \emph{259},
  114096\relax
\mciteBstWouldAddEndPuncttrue
\mciteSetBstMidEndSepPunct{\mcitedefaultmidpunct}
{\mcitedefaultendpunct}{\mcitedefaultseppunct}\relax
\EndOfBibitem
\bibitem[Langston \latin{et~al.}(2025)Langston, Ruiz, Boix-Constant,
  Ma{\~n}as-Valero, Coronado, Baldov{\'\i}, Liu, and
  Musfeldt]{langston2025pressure}
Langston,~L.~J.; Ruiz,~A.~M.; Boix-Constant,~C.; Ma{\~n}as-Valero,~S.;
  Coronado,~E.; Baldov{\'\i},~J.~J.; Liu,~Z.; Musfeldt,~J.~L. Pressure-induced
  structural phase transitions in CrSBr. \emph{npj Quantum Materials}
  \textbf{2025}, \emph{10}, 57\relax
\mciteBstWouldAddEndPuncttrue
\mciteSetBstMidEndSepPunct{\mcitedefaultmidpunct}
{\mcitedefaultendpunct}{\mcitedefaultseppunct}\relax
\EndOfBibitem
\bibitem[Bykov \latin{et~al.}(2013)Bykov, Bykova, van Smaalen, Dubrovinsky,
  McCammon, Prakapenka, and Liermann]{Bykov2013}
Bykov,~M.; Bykova,~E.; van Smaalen,~S.; Dubrovinsky,~L.; McCammon,~C.;
  Prakapenka,~V.; Liermann,~H.-P. High-pressure behavior of FeOCl.
  \emph{Physical Review B} \textbf{2013}, \emph{88}\relax
\mciteBstWouldAddEndPuncttrue
\mciteSetBstMidEndSepPunct{\mcitedefaultmidpunct}
{\mcitedefaultendpunct}{\mcitedefaultseppunct}\relax
\EndOfBibitem
\bibitem[Schaller \latin{et~al.}(2022)Schaller, Bykov, Bykova, Glazyrin, and
  van Smaalen]{Schaller2022}
Schaller,~A.~M.; Bykov,~M.; Bykova,~E.; Glazyrin,~K.; van Smaalen,~S.
  Pressurizing the van der Waals magnet FeOCl at low temperatures: Phase
  transitions and structural evolution. \emph{Physical Review B} \textbf{2022},
  \emph{105}\relax
\mciteBstWouldAddEndPuncttrue
\mciteSetBstMidEndSepPunct{\mcitedefaultmidpunct}
{\mcitedefaultendpunct}{\mcitedefaultseppunct}\relax
\EndOfBibitem
\bibitem[Schaller \latin{et~al.}(2023)Schaller, Bykov, Bykova, Glazyrin, Rekis,
  and van Smaalen]{Schaller2023}
Schaller,~A.~M.; Bykov,~M.; Bykova,~E.; Glazyrin,~K.; Rekis,~T.; van
  Smaalen,~S. Pressure-dependent distortions in quasi-two-dimensional magnetic
  CrOCl at low temperatures. \emph{Physical Review B} \textbf{2023},
  \emph{108}\relax
\mciteBstWouldAddEndPuncttrue
\mciteSetBstMidEndSepPunct{\mcitedefaultmidpunct}
{\mcitedefaultendpunct}{\mcitedefaultseppunct}\relax
\EndOfBibitem
\bibitem[Kotla \latin{et~al.}(2025)Kotla, Ramakrishnan, Schaller, Rekis,
  Eisele, Bao, Noohinejad, {de Laitre}, {de Boissieu}, and {van
  Smaalen}]{kotla2025a}
Kotla,~S.~R.; Ramakrishnan,~S.; Schaller,~A.~M.; Rekis,~T.; Eisele,~C.;
  Bao,~J.-K.; Noohinejad,~L.; {de Laitre},~G.; {de Boissieu},~M.; {van
  Smaalen},~S. Deciphering the commensurately modulated monoclinic phase of
  {Rb$_2$ZnCl$_4$} at low temperatures. \emph{Journal of Solid State Chemistry}
  \textbf{2025}, \emph{345}, 125226\relax
\mciteBstWouldAddEndPuncttrue
\mciteSetBstMidEndSepPunct{\mcitedefaultmidpunct}
{\mcitedefaultendpunct}{\mcitedefaultseppunct}\relax
\EndOfBibitem
\bibitem[Chapuis \latin{et~al.}(1980)Chapuis, Hulliger, and
  Schmelczer]{chapuis1980a}
Chapuis,~G.; Hulliger,~F.; Schmelczer,~R. The crystal structure and some
  properties of Eu2Sb3. \emph{Journal of Solid State Chemistry} \textbf{1980},
  \emph{31}, 59--67\relax
\mciteBstWouldAddEndPuncttrue
\mciteSetBstMidEndSepPunct{\mcitedefaultmidpunct}
{\mcitedefaultendpunct}{\mcitedefaultseppunct}\relax
\EndOfBibitem
\bibitem[Ramakrishnan \latin{et~al.}(2026)Ramakrishnan, Gao, Olevano, Pachoud,
  Hadj-Azzem, Gabarino, Perez, Pautrat, Valenti, Quenot, Pairis, Chernyshov,
  Noohinejad, Paulmann, Bulled, Bosak, van Smaalen, Toulemonde, Measson, and
  Rodiere]{ramakrishnan2026a}
Ramakrishnan,~S. \latin{et~al.}  Tracing the horizon of
  tetragonal-to-monoclinic distortion in pressurized trilayer nickelate
  La4Ni3O10. \emph{Communications Materials} \textbf{2026}, \emph{7}, 248\relax
\mciteBstWouldAddEndPuncttrue
\mciteSetBstMidEndSepPunct{\mcitedefaultmidpunct}
{\mcitedefaultendpunct}{\mcitedefaultseppunct}\relax
\EndOfBibitem
\bibitem[Hill(2000)]{Hill2000}
Hill,~N.~A. Why Are There so Few Magnetic Ferroelectrics? \emph{The Journal of
  Physical Chemistry B} \textbf{2000}, \emph{104}, 6694–6709\relax
\mciteBstWouldAddEndPuncttrue
\mciteSetBstMidEndSepPunct{\mcitedefaultmidpunct}
{\mcitedefaultendpunct}{\mcitedefaultseppunct}\relax
\EndOfBibitem
\bibitem[Mondal \latin{et~al.}(2025)Mondal, Markina, Hopf, Krelle, Shradha,
  Klein, Glazov, Gerber, Hagmann, Klitzing, Mosina, Sofer, and
  Urbaszek]{Mondal2025}
Mondal,~P.; Markina,~D.~I.; Hopf,~L.; Krelle,~L.; Shradha,~S.; Klein,~J.;
  Glazov,~M.~M.; Gerber,~I.; Hagmann,~K.; Klitzing,~R.~v.; Mosina,~K.;
  Sofer,~Z.; Urbaszek,~B. Raman polarization switching in CrSBr. \emph{npj 2D
  Materials and Applications} \textbf{2025}, \emph{9}\relax
\mciteBstWouldAddEndPuncttrue
\mciteSetBstMidEndSepPunct{\mcitedefaultmidpunct}
{\mcitedefaultendpunct}{\mcitedefaultseppunct}\relax
\EndOfBibitem
\bibitem[Sahu \latin{et~al.}(2025)Sahu, Berrezueta-Palacios, Juergensen,
  Mosina, Sofer, Velický, Kusch, and Frank]{Sahu2025}
Sahu,~S.; Berrezueta-Palacios,~C.; Juergensen,~S.; Mosina,~K.; Sofer,~Z.;
  Velický,~M.; Kusch,~P.; Frank,~O. Resonance Raman scattering and anomalous
  anti-Stokes phenomena in CrSBr. \emph{Nanoscale} \textbf{2025}, \emph{17},
  11539–11546\relax
\mciteBstWouldAddEndPuncttrue
\mciteSetBstMidEndSepPunct{\mcitedefaultmidpunct}
{\mcitedefaultendpunct}{\mcitedefaultseppunct}\relax
\EndOfBibitem
\bibitem[Long \latin{et~al.}(2023)Long, Mosina, H{\"u}bner, Sofer, Klein,
  Prucnal, Helm, Dirnberger, and Zhou]{long2023intrinsic}
Long,~F.; Mosina,~K.; H{\"u}bner,~R.; Sofer,~Z.; Klein,~J.; Prucnal,~S.;
  Helm,~M.; Dirnberger,~F.; Zhou,~S. Intrinsic magnetic properties of the
  layered antiferromagnet CrSBr. \emph{Applied Physics Letters} \textbf{2023},
  \emph{123}\relax
\mciteBstWouldAddEndPuncttrue
\mciteSetBstMidEndSepPunct{\mcitedefaultmidpunct}
{\mcitedefaultendpunct}{\mcitedefaultseppunct}\relax
\EndOfBibitem
\bibitem[Rigaku(2023)]{crysalis}
Rigaku {CrysAlis Pro Version 171.43.90, Rigaku Oxford Diffraction.} 2023\relax
\mciteBstWouldAddEndPuncttrue
\mciteSetBstMidEndSepPunct{\mcitedefaultmidpunct}
{\mcitedefaultendpunct}{\mcitedefaultseppunct}\relax
\EndOfBibitem
\bibitem[Shen \latin{et~al.}(2020)Shen, Wang, Dewaele, Wu, Fratanduono, Eggert,
  Klotz, Dziubek, Loubeyre, Fat’yanov, Asimow, Mashimo, and
  Wentzcovitch]{Shen2020}
Shen,~G.; Wang,~Y.; Dewaele,~A.; Wu,~C.; Fratanduono,~D.~E.; Eggert,~J.;
  Klotz,~S.; Dziubek,~K.~F.; Loubeyre,~P.; Fat’yanov,~O.~V.; Asimow,~P.~D.;
  Mashimo,~T.; Wentzcovitch,~R. M.~M. Toward an international practical
  pressure scale: A proposal for an IPPS ruby gauge (IPPS-Ruby2020). \emph{High
  Pressure Research} \textbf{2020}, \emph{40}, 299--314\relax
\mciteBstWouldAddEndPuncttrue
\mciteSetBstMidEndSepPunct{\mcitedefaultmidpunct}
{\mcitedefaultendpunct}{\mcitedefaultseppunct}\relax
\EndOfBibitem
\bibitem[Breslavetz \latin{et~al.}(2021)Breslavetz, Delhomme, Pelini, Pawbake,
  Vaclavkova, Orlita, Potemski, Measson, and Faugeras]{breslavetz2021spatially}
Breslavetz,~I.; Delhomme,~A.; Pelini,~T.; Pawbake,~A.; Vaclavkova,~D.;
  Orlita,~M.; Potemski,~M.; Measson,~M.-A.; Faugeras,~C. Spatially resolved
  optical spectroscopy in extreme environment of low temperature, high magnetic
  fields and high pressure. \emph{Review of Scientific Instruments}
  \textbf{2021}, \emph{92}\relax
\mciteBstWouldAddEndPuncttrue
\mciteSetBstMidEndSepPunct{\mcitedefaultmidpunct}
{\mcitedefaultendpunct}{\mcitedefaultseppunct}\relax
\EndOfBibitem
\bibitem[Meyer \latin{et~al.}()Meyer, Elsässer, and M.]{meyer}
Meyer,~B.; Elsässer,~C.; M.,~F. FORTRAN90 program for mixed-basis calculations
  for crystals. \emph{Max-Planck-Institut für Metallforschung, Stuttgart}
  \relax
\mciteBstWouldAddEndPunctfalse
\mciteSetBstMidEndSepPunct{\mcitedefaultmidpunct}
{}{\mcitedefaultseppunct}\relax
\EndOfBibitem
\bibitem[Perdew \latin{et~al.}(1996)Perdew, Burke, and Ernzerhof]{pbe1996}
Perdew,~J.~P.; Burke,~K.; Ernzerhof,~M. Generalized gradient approximation made
  simple. \emph{Phys. Rev. Lett.} \textbf{1996}, \emph{77}, 3865\relax
\mciteBstWouldAddEndPuncttrue
\mciteSetBstMidEndSepPunct{\mcitedefaultmidpunct}
{\mcitedefaultendpunct}{\mcitedefaultseppunct}\relax
\EndOfBibitem
\bibitem[Heid and Bohnen(1999)Heid, and Bohnen]{heid1999}
Heid,~R.; Bohnen,~K.~P. Linear response in a density-functional mixed-basis
  approach. \emph{Phys. Rev. B} \textbf{1999}, \emph{60}, R3709 (R)\relax
\mciteBstWouldAddEndPuncttrue
\mciteSetBstMidEndSepPunct{\mcitedefaultmidpunct}
{\mcitedefaultendpunct}{\mcitedefaultseppunct}\relax
\EndOfBibitem
\end{mcitethebibliography}

\providecommand{\latin}[1]{#1}
\makeatletter
\providecommand{\doi}
  {\begingroup\let\do\@makeother\dospecials
  \catcode`\{=1 \catcode`\}=2 \doi@aux}
\providecommand{\doi@aux}[1]{\endgroup\texttt{#1}}
\makeatother
\providecommand*\mcitethebibliography{\thebibliography}
\csname @ifundefined\endcsname{endmcitethebibliography}
  {\let\endmcitethebibliography\endthebibliography}{}

\end{document}